# Preparing graduate students to be educators

Edward Price[a]
Department of Physics
California State University, San Marcos
San Marcos, CA 92096

Noah Finkelstein[b]
Department of Physics
University of Colorado at Boulder
Boulder, Colorado 80309

**Abstract**
We present two programs that address needs to better prepare graduate students for their roles as professional physicists, particularly in the areas of teaching and education research. The two programs, Preparing Future Physicists (PFP) and a course, *Teaching and Learning Physics,* are designed to be mutually supportive, address these broader graduate roles, and support the development of the field of physics education research. While voluntary, PFP has attracted the participation of roughly half the physics graduate students at each of two large research institutions. Compared to the national rate, these students are roughly twice as likely to report an interest in pursuing future roles as educators. While less than one in five of participants surveyed reported education being valued by the research community in physics, more than 90% reported intentions to incorporate the results of research in physics education in their future teaching. Experience with the synergistic program, *Teaching and Learning Physics*, demonstrates that it is possible to replicate earlier successes of the program initiated at a different institution, including increasing student mastery of physics, developing student interest in education and teaching, and engaging students in research projects in physics education. In addition to introducing these programs, we identify some of the critical features that contribute to their successes.

# I. Introduction

It is now well recognized that we need to better educate our undergraduates in physics and the sciences more broadly.[1-6] As a result of research, we know of curricula, procedures, and approaches that effectively educate our undergraduates.[7-9] Yet, we do not broadly prepare our future faculty to develop or to implement these now well-understood research-based educational practices. Though professional development can have an important impact on current faculty, here we focus on graduate school as a critical experience in the preparation of future physicists, and as a key point of leverage to integrate teaching and education research into the broader physics culture.

The development of graduate student research skills tend to follow a studied and productive framework; in contrast, the development of teaching skills and training in education follow a more *ad hoc*, or folk-theory of preparation. During graduate school, physicists engage in authentic research experiences, but there is no corresponding apprenticeship regarding teaching and learning.[10, 11] Despite the need for institutions that emphasize research, we note that the minority of graduate students become faculty at institutions similar to those in which they were trained,[10, 12-15] and all students will benefit from a broader preparation. A recent report on the state of physics graduate education recommends better informing graduates about the full range of employment opportunities for graduate students, calls for the development of graduate students' communication skills, and encourages innovative methods for the delivery of the graduate curriculum.[16] We present two programs that address these recommendations, particularly through the attention to education and teaching in graduate preparation.

Addressing the needs for improved undergraduate education and broader graduate education can also support and strengthen the ongoing growth of the field of physics education research, PER. By extending the focus of physics graduate school to include structured attention to education, we begin to give education greater prominence and validate education research and reform in physics, by physicists. In this way, we may broadly educate our graduate students, shift the culture of physics to include education in the core practice of physicists, and sustain and extend the work of physics education researchers.

There are excellent model programs designed to support the development of physics graduate students as educators and professionals in physics.[17-19] In this paper, we examine two programs at University of Colorado (CU) and University of California, San Diego (UCSD) that are designed to couple with each other and to address broader graduate roles and support the development of the field of physics education research. In addition to introducing these programs and documenting their successes, we identify critical features that make them successful. These programs focus on student participation (generally voluntary), have tiered levels of involvement and commensurate levels of commitment, engage students in reflective practice, provide students with practical experiences teaching and researching in education, and build a community of scholars committed to the inclusion of education in the practice of physicists.

## II. Preparing Future Physicists – A graduate development program
### A. Program background
In 1998, the American Association of Physics Teachers (AAPT) funded Preparing Future Physics Faculty (PFPF), a graduate program designed to augment traditional training in research. PFPF was a discipline-specific version of Preparing Future Faculty, a program initiated by the Council of Graduate Schools and the Association of American Colleges and Universities.[17] PFPF and PFF were responses to calls for increased emphasis on preparation in the areas of teaching and professional development by the Association of American Universities and the National Academy of Sciences.[20, 21] UCSD was one of the sites chosen for a PFPF program. One of the authors [NF] was involved with establishing the program; the other [EP] is a former participant and director. The program ran with external support until 2000, and has since continued with the support of the physics department and UCSD's campus-wide Center for Teaching Development.

### B. Program description
Initially, the PFF/PFPF program was intended to reshape graduate preparation to "produce students who are well prepared to meet the needs of institutions that hire new faculty" by including an emphasis on teaching and professional development.[17] Over the eight years of its existence, the UCSD instantiation of the program has undergone substantial changes and evolved to address four goals:
- Preparing graduate students for their future responsibilities as educators by promoting awareness and understanding of PER;
- Raising awareness of differences in the needs and opportunities at different academic institutions (i.e., community colleges, bachelor's granting institutions, and regional and research universities);
- Providing physics graduate students with professional and career development in areas such as conducting a job search and writing grant proposals;
- Creating an environment where physics graduate students discuss issues in the physics community.

Participation in the program is voluntary and can include a variety of activities, depending on the students' interests and time. At a minimum, participants attend weekly (or bi-weekly) seminars on topics relating to the goals discussed above. Table I indicates how particular seminars address these themes. Given the shift from the program's original focus, and in light of the participants' broad career interests (discussed below), we are now calling the program Preparing Future Physicists (PFP).

In addition to weekly seminars, graduate students are encouraged to participate in a range of practice-based activities: researching, developing curricula, and teaching; examples are shown in Figure 1. Research projects include graduate students engaging in PER-based studies of local practice (such as examining instructor beliefs about teaching). Curricular development often takes the form of graduate students appropriating PER-based activities and adopting them for local practice. For example, graduate students have augmented the complement of Interactive Lecture Demonstrations[22] (ILDs) running in the introductory physics sequence by building an RC circuit ILD (and testing its effectiveness in the algebra-based course). Lastly, teaching practice is heavily emphasized. All students are encouraged to conduct a 5-10 minute micro-teach (presenting a single topic to the rest of the PFP seminar). Subsequently, students engage in observations and guest lectures in local introductory courses and at partner institutions

(community and teaching colleges). Ultimately, several students have become instructors-of-record, taking responsibility for designing and implementing a full term class at these partner institutions. All of these activities are overseen both locally by the PFP organizer and at the host institutions by practicing faculty. These research and teaching activities ground the seminar discussions in practical experience, making both more meaningful.

---

Weekly seminar:

*Microteach:* The seminar begins with a participant giving a brief (5min) demonstration or explanation to the whole group, followed by feedback from the other participants.

*Discussion of interactive lecture demonstrations (ILDs):* Participants experience an ILD as 'students', quickly leading to discussions of friction and sampling rate. The discussion is refocused on analyzing the experience and the use of ILDs in large lecture courses. Evidence for the effectiveness of ILDs is discussed, and the ILD process is connected to prior discussions of theories of learning and teaching. Participants are given copies of relevant articles in the literature.

Practice-based activity:

*Curriculum development project:* Two participants develop an RC flasher circuit interactive demonstration, employing real time data acquisition and the ILD protocol. Subsequent refinements are based on feedback during a microteach and field-testing with students in large lecture course.

---

**Figure 1.** Example PFP activities.

The scope of graduate student engagement (ranging from participation in seminar discussions, to curricular development, to teaching a course as instructor-of-record) depends on the participant's interests and constraints. Guest lecturing is valuable experience with a small time commitment. Conversely, teaching a course provides a more comprehensive experience but is a demanding undertaking. Our most successful participant activities combine the best of both approaches by including a group planning component and a modular workload. By involving multiple participants, these programs can have a significant impact, while only requiring modest effort from individual graduate students. In general, the PFP program is structured so that varied levels of participation are legitimate, and students are encouraged to participate at a level they find appropriate.

**C. Sustainability and scalability**

The program's tiered-participation model has been remarkably robust through several changes in program leadership. We attribute this to a few essential features: the involvement of a program organizer, sustained graduate student interest in the issues addressed by the program, recognition of this program as legitimate within physics, and a flexible format that allows the program to reflect the participants' and organizer's interests. Except for modest funding, official administrative support has not been essential, and in fact has lagged behind the bottom-up support for the program. (Three years ago, participation in the program was officially recognized as fulfilling the UCSD physics department's teaching requirement; this year, for the first time, the department officially recognized the organizer's effort by granting teaching relief.) Following the initial framework developed at UCSD, NF implemented a PFP program two years ago at the University of Colorado.[23] The model's central framing – voluntary participation of graduate students in tiered levels of participation – has remained the same.  More on the UCSD program can be found at http://www.ctd.ucsd.edu/programs/pfpf/ and the CU program at http://per.colorado.edu/pfpf

## III. Teaching and Learning Physics – A graduate course in education research and practice

### A. Course background and description

Complementing the PFP program, we present another model for incorporating educational issues in graduate preparation – a course in teaching and learning physics that provides an intensive focus on physics education and physics education research. Intended for graduate students interested in the study of education, the course is formalized institutionally through course credit. Initially developed by NF in 1998 at UCSD and subsequently implemented in 2003 at CU, the physics course *Teaching and Learning Physics* is structured around three central components: study of pedagogical issues (cognitive, psychological, educational), study of physics content, and practical experience teaching in the community (both in local community and within the University).  Each of these course components complements the others by providing a differing perspective on the same area of inquiry.  For example, the same week that students read studies documenting individuals' difficulties with the electric field, the students study the concept itself, and teach it to others.  This model has been described in detail previously.[24] This course attracts students to physics from all demographic backgrounds, increases the number of physics majors enrolling in teacher education, and builds strong and sustainable ties between the university and community partners. The course model has been employed elsewhere, with colleagues conducting versions of this course in at least five different research institutions. One such example is described in the accompanying paper by Wittmann and Thompson in this issue.

Particularly relevant, the course on teaching and learning physics engages students in research activities throughout – applying tools of science to education.  Students' projects in the course allow them to view the practices of education, teaching, and learning as scholarly pursuits.  The resultant projects have spanned from developing after-school programs that increase younger students' interest and acuity in physics, to programs that study the role of gender in the university classroom.  Several of these projects have led to published work (for instance see references 25, 26), while others have led to the creation of community partnerships that would not have otherwise existed (such as the CU STOMP program[27] or UCSD's Fleet University[28]). Other student research and teaching efforts have been instrumental in the implementation of educational reforms at the university.  For example, at CU, the implementation of *Tutorials in Introductory Physics*[29] in the undergraduate calculus based introductory sequence required an increased teacher-to-student ratio. Students from the course on teaching and learning physics provided critical human resources,[30] while the *Tutorials* provided real world examples of educational reforms that graduate students could study. Each of these activities provides students the opportunity to engage in authentic educational practices, while also sending the message that these activities are part of a physicist's pursuits.

## IV. Outcomes and discussion

### A. Student participation

In the broadest sense, PFP and *Teaching and Learning Physics* represent attempts to more thoroughly prepare graduate students by more fully including education in the core practice of physicists. While it should be clear that affecting students' choices and preparation is a long-term endeavor, we may assess the preliminary impact of these programs. First, it is worth considering whether students choose to participate in these voluntary programs. In the graduate program, participation has increased since its inception; starting with fewer than ten students, the UCSD program now regularly supports about twenty students. In a given year, there are about 115 physics graduate students at UCSD, so that over five to six years of graduate studies, a student is about as likely to participate in PFP as not. In the CU version of PFP, average attendance is roughly thirty graduate students per session and over 100 individuals have participated in the last year. In the course, *Teaching and Learning Physics*, ten to fifteen students have participated annually since its inception at UCSD, and in its first offering at CU, 23 students enrolled. Graduate students are clearly interested in engaging the issues addressed in these programs, and there are few other outlets for this interest.[10]

**B. Student interests and career goals**

To gauge graduate students' career interests and motivations for participating in the PFP program, current UCSD and CU PFP participants were informally surveyed during spring 2006. Table II indicates student responses to our questionnaire. For comparison, the table also includes data from Statistical Research Center of the American Institute of Physics on the top choice of career path for US physics graduate students.[31] Similar percentages of respondents from each group selected 'Research tenure-track faculty' as their first choice. But, PFP respondents chose 'Faculty at liberal arts or community college or other teaching' at more than twice the rate of US physics graduate students overall, while choosing 'Research in industry, national lab or university setting' at less than half the rate of US physics graduate students overall. Based on these responses, PFP participants are more interested in academic careers and teaching than typical physics graduate students.[32] When asked what they "really think they'll be doing 10 years from now," the percent of PFP respondents choosing 'Research tenure-track faculty' dropped to 16%, while the percent choosing 'Faculty at liberal arts or community college or other teaching' increased to 44%. Participants were also asked what the PFP program would, ideally, do for them. Reflecting the participants' career hopes and expectations, 34% of respondents selected 'provide a diversity of ideas for my future' as their top choice, and 27% of respondents selected 'provide ideas for being a more effective educator'.

**C. Participant outcomes**

Building bridges between physics and education and infusing physics education research into traditional practices in physics are central goals of these programs. As measured by other surveys of the participants, each of these programs has been successful at addressing these interrelated goals. We have surveyed PFP participants on their attitudes about the importance of teaching and what they have learned from the program.[33] Results from 2001-2002, 2002-2003, and 2003-2004 cohorts at UCSD are summarized in Table III. While few respondents feel education is valued by the physics research community, most plan on incorporating the results of PER in their own teaching. Informal contacts with former participants who are now teaching suggest that they are following through on these intentions. Finally, when asked if they were considering entering the education research field, less than 40% of respondents agreed, suggesting that the program is in fact reaching beyond the PER community by serving the dual purposes of supporting those students interested in PER and those students who may simply employ the research findings of PER in their own teaching practices. Students enrolled in the course on teaching and learning physics report it to be among their most favored and useful courses. Table IV shows student responses to select items on an end of semester course evaluation. Students in the course engage deeply and look for more learning opportunities.

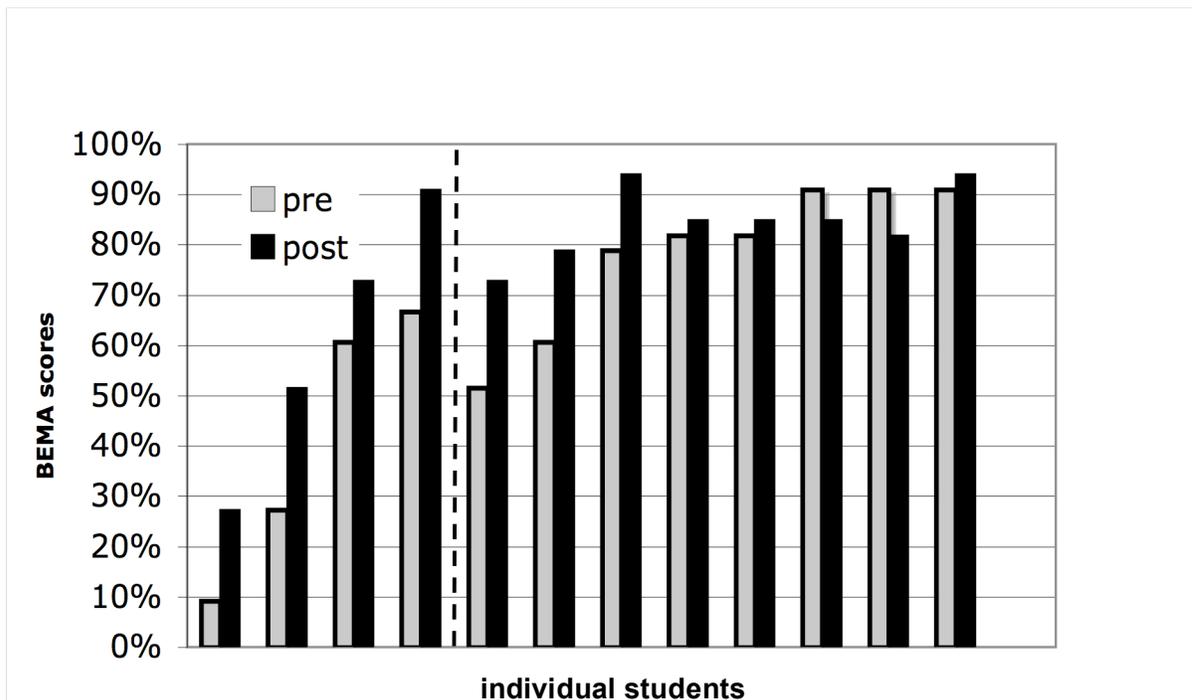

**Figure 2.** Pre/post conceptual EM assessment of CU Teaching and Learning Physics

In its initial instantiation at UCSD, *Teaching and Learning Physics* was demonstrated to increase student mastery of physics, proficiency at teaching, and the likelihood that students engage in future teaching experiences.[24] Furthermore, evaluation of student understanding of education reveal a shift from a more transmissionist perspective to a more progressive, constructivist perspective.[24] More recent, corroborating, results have been obtained at CU. As indicated in Figure 2, students in the course showed significant normalized gain on the Brief Electricity and Magnetism Assessment (BEMA), an assessment of conceptual understanding of electricity and magnetism.[34] The vertical dashed line in the figure separates scores for undergraduate (to the left) and graduate (to the right) students. The course normalized gain is 38%,[35] and a two-tailed, paired t-test on the pre- and post-test scores indicates statistically significant differences of $p<0.02$. Those students who participated in more formalized teaching roles, as Learning Assistants[36] in the Colorado implementation of Tutorials in Introductory Physics,[30] posted normalized learning gains of 55%, while their classmates who taught or conducted research in other environments posted normalized learning gains of 27%, with $p<0.1$ for a two-tailed t-test comparing the two groups. These gains are striking, given that physics content is only one component of the course, and all students had taken between one and three courses in E/M previously.

**D. Broader outcomes and synergies**
In addition to impacts on the graduate student participants, PFP and *Teaching and Learning Physics* have had broader, positive impacts. The programs create a pool of graduate students that are well prepared for teaching positions within the department. Furthermore, participants form a network that can assist and support the implementation of research-based teaching methods. Fieldwork and projects, undertaken through participation in PFP and *Teaching and Learning Physics*, often support outreach programs in the broader community. PFP has provided a mechanism for strengthening interactions between the host physics departments and other local institutions, such as teaching-focused colleges and informal science centers. These interactions include seminars by guests from partner institutions, PFP participants teaching at partner institutions, and informal activities such as visits to partner institutions. These interactions provide important exposure to varied institutional environments. For many participants these contacts may be the first time they have heard someone say that teaching physics at a community college is a satisfying and rewarding career. This community network is an important aspect of the program, given participants' career interests and reasons for participating.

As a final outcome, we consider interactions between the two activities. Though the PFP program and the course *Teaching and Learning Physics* are independent and modular, they form a mutually-supportive continuum of increasing level of engagement, with related, but distinct, focuses. As a result, students interested in physics education or physics education research can participate with an emphasis and intensity that they find appropriate. Interactions between the programs lead to benefits for both; for instance, PFP creates a pool of students interested in further study, while *Teaching and Learning Physics* creates 'expert' participants that enrich PFP discussions and activities. Institutional support is developed from broad student interest, the value of the programs' "products" (curriculum development, instructional reform), and the benefits to the graduate participants.

### E. Key features

In our analysis of these programs, we have identified five features that contribute to the programs' success and durability. First, students are active participants who help construct the program, not merely observers, and participation is generally voluntary. Second, the programs provide tiered levels of participation, allowing students to choose a degree of engagement they find appropriate, and that may change over time as their interests and commitments evolve. Third, participants can engage in direct, practical experiences in teaching and in education research. Fourth, the programs provide a framework for reflection, allowing participants to reconcile formal ideas with personal experiences. Finally, the programs build and support a community of physicists with shared commitments to education.

## V. Conclusion

We have described two activities designed to broaden physics graduate students' conception of and preparation for their profession by focusing on education and education research. These efforts are part of a broader goal of including education as an essential part of what it means to be a physicist. This broad goal is compatible with recent calls to improve undergraduate science teaching and graduate preparation, and supports the continued growth of PER and the adoption of PER-based teaching methods. Through participation in these programs, graduate students come to value education more deeply as a core practice of physicists. More broadly, these programs can lead to similar shifts in local culture. While it is not certain that these shifts will be sustained, by creating layered and complementary programs these changes are more robust. Though many graduate program reforms have the intent of changing the preparation of graduate students in order to support the changing job market, it may turn out that graduate students involved in the programs described above will change the nature of the discipline.

## Acknowledgments

This paper is based on an article in the Spring 2006 APS Forum on Education Newsletter. We are grateful for valuable input and support from Barbara Jones, Omar Clay, Rosalind Streichler, Tom Murphy, Christopher Keller, the graduate participants in the PFP programs, and the students in *Teaching and Learning Physics*. Thanks to Rachael Scherr for valuable feedback on this paper. We would also like to acknowledge the NSF and AAPT's initial funding of the PFPF program, the continuing support of the UCSD and CU physics departments, and the UCSD Center for Teaching Development.

**Table I: Areas and topics in Preparing Future Physicists program.**

| Physics Education | Professional & career development | Diverse academic environments | The physics community |
|---|---|---|---|
| • Current state of physics education<br>• Cognitive issues in learning physics<br>• Conceptual-question performance of intro physics classes<br>• Physics demonstrations<br>• Interactive Lecture Demos<br>• Peer instruction & class response systems<br>• Effective presentation & lecturing<br>• Educational technology<br>• Assessment: The role of testing & classes of questions | • Choosing a graduate advisor<br>• The departmental examination<br>• Postdoctoral positions: A panel discussion with current postdocs<br>• The academic job market in physics<br>• Resume/CV & application material prep + interviewing<br>• Grant writing panel & grad fellowship opportunities | • Physics in regional universities<br>• Physics in liberal arts colleges<br>• Physics in high schools<br>• Physics in community colleges | • Political & social issues in physics<br>• Gender Issues in physics<br>• Corporate – university interactions<br>• Physics in industry/government labs |

**Table II: Spring 2006 survey of UCSD and CU PFP participants.**

| Question | PFP | US grads[a] |
|---|---|---|
| **What do you really hope to be doing 10 years from now?** | | |
| *Research tenure-track faculty* | 38% | 41%[b] |
| *Faculty at liberal arts or community college or other teaching* | 31 | 13[c] |
| *Research in industry, national lab or university setting* | 15 | 37[d] |
| *Private consulting or other self-directed venture* | 4 | 3[e] |
| *Something not on this list* | 12 | 6[f] |
| **What do you really think you're going to be doing in 10 years?** | | |
| *Research tenure-track faculty* | 16 | |
| *Faculty at liberal arts or community college or other teaching* | 44 | |
| *Research in industry, national lab or university setting* | 20 | |
| *Private consulting or other self-directed venture* | 4 | |
| *Something not on this list* | 16 | |
| **What would PFPF ideally do for you (besides feed you pizza)?** | | |
| *Provide a diversity of ideas for my future* | 34 | |
| *Help me plot a course to faculty-hood* | 12 | |
| *Provide ideas for being a more effective educator* | 27 | |
| *Bring awareness of social issues that physicists can address* | 19 | |
| *Teach me the skills I will need as a faculty-person* | 8 | |

[a]Based on AIP Statistical Research Center's 2000-01 Survey of Graduate Physics and Astronomy Students, Table 11, which reports 'top choice of career path for PhD level, US citizen, physics students'. AIP response categories were: [b]research or teach at a university; [c]research or teach at a four year college; [d]work in an industrial research and development setting, **or** work in the field of information systems and computers for a private company, **or** work for government or a national lab; [e]be self-employed or consultant; [f]all other.

**Table III: Survey of UCSD PFP Participants.**

| Of the 33/38 respondents who considered education a substantive part of their future career: | |
|---|---|
| | **Percent agreeing** |
| I feel education is valued by the physics research community | 18% |
| **After participation in PFP…** | |
| I am more aware of the results of PER | 82% |
| I am planning to incorporate PER results in teaching | 94% |
| I view PER as a legitimate research activity with in physics community | 88% |

**Table IV: Teaching and Learning Physics course evaluations by 18 CU students in Spring 2006.**

| | *Extremely* | *Somewhat* | *Not at all* |
|---|---|---|---|
| *How useful to me is this class:* | **89%** | **11%** | **0%** |
| *How enjoyable is this class:* | **94%** | **6%** | **0%** |
| | *A great deal* | *Something* | *Nothing at all* |
| *How much did you learn:* | **94%** | **6%** | **0%** |
| | *Enthusiastically* | *Maybe* | *Never* |
| *I recommend this course to others:* | **100%** | **0%** | **0%** |
| | *They must* | *Maybe* | *Definitely not* |
| *The department should offer this course in the future:* | **100%** | **0%** | **0%** |